\documentclass[%
longbibliography,
 reprint,
 showkeys,
superscriptaddress,
 amsmath,amssymb,
 pre,
]{revtex4-1}

\usepackage{graphicx}% Include figure files
\usepackage{dcolumn}% Align table columns on decimal point
\usepackage{bm}% bold math
\usepackage[utf8]{inputenc}
\usepackage{graphicx}
\usepackage{color}
\usepackage{amsmath}
\usepackage{amssymb}
\usepackage{subfigure}
\usepackage{epsfig}
\usepackage{float}
\usepackage{lipsum}
\usepackage{cases}
\usepackage{appendix}
\usepackage{soul}

\newcommand{\D}{\boldsymbol{D}}

\newcommand{\M}{\boldsymbol{M}}

\newcommand{\B}{\boldsymbol{B}}

\newcommand{\nn}{\boldsymbol{n}}
\newcommand{\J}{\boldsymbol{J}}

\newcommand{\vv}{\boldsymbol{v}}
\newcommand{\rr}{\boldsymbol{r}}
\newcommand{\tet}{\boldsymbol{\theta}}

\newcommand{\0}{\boldsymbol{0}}
\newcommand{\ov}{\boldsymbol{s}}

\newcommand{\eeta}{\boldsymbol{\eta}}

\newcommand{\irow}[1]{% inline row vector
  \begin{matrix}(#1)\end{matrix}%
}

\begin{document}

%\preprint{APS/123-QED}

%\articletype{ARTICLE TEMPLATE}% Specify the article type or omit as appropriate

\title{Unusual stationary state in Brownian systems with Lorentz force}

\author{I.~Abdoli}
\affiliation{Leibniz-Institut  f\"ur Polymerforschung Dresden, Institut Theorie der Polymere, 01069 Dresden, Germany}

\author{H.D.~Vuijk}
\affiliation{Leibniz-Institut  f\"ur Polymerforschung Dresden, Institut Theorie der Polymere, 01069 Dresden, Germany}

\author{R.~Wittmann}
\affiliation{Institut f\"ur Theoretische Physik II, Weiche Materie, Heinrich-Heine-Universit\"at D\"usseldorf, 40225 D\"usseldorf, Germany}

%\author{H.~Merlitz}
%\affiliation{Leibniz-Institut  f\"ur Polymerforschung Dresden, Institut Theorie der Polymere, 01069 Dresden, Deutschland}

\author{J.U.~Sommer}
\affiliation{Leibniz-Institut  f\"ur Polymerforschung Dresden, Institut Theorie der Polymere, 01069 Dresden, Germany} \affiliation{Technische Universit\"at Dresden, Institut f\"ur Theoretische Physik, 01069 Dresden, Germany}

\author{J.M.~Brader}
\affiliation{Department de Physique, Universit\'e de Fribourg, CH-1700 Fribourg, Switzerland}

\author{A.~Sharma}
\email{sharma@ipfdd.de}
\affiliation{Leibniz-Institut  f\"ur Polymerforschung Dresden, Institut Theorie der Polymere, 01069 Dresden, Germany} \affiliation{Technische Universit\"at Dresden, Institut f\"ur Theoretische Physik, 01069 Dresden, Germany}

\begin{abstract}
In systems with overdamped dynamics, the Lorentz force reduces the diffusivity of a Brownian particle in the plane perpendicular to the magnetic field. The anisotropy in diffusion implies that the Fokker-Planck equation for the probabiliy distribution of the particle acquires a tensorial coefficient.
The tensor, however, is not a typical diffusion tensor due to the antisymmetric elements which account for the fact that Lorentz force curves the trajectory of a moving charged particle. This gives rise to unusual dynamics with features such as additional Lorentz fluxes and a nontrivial density distribution, unlike a diffusive system. The equilibrium properties are, however, unaffected by the Lorentz force. Here we show that by stochastically resetting the Brownian particle, a nonequilibrium steady state can be created which preserves the hallmark features of dynamics under Lorentz force. We then consider a minimalistic example of spatially inhomogeneous magnetic field, which shows how Lorentz fluxes fundamentally alter the boundary conditions giving rise to an unusual stationary state.

\end{abstract}

%\keywords{
%Brownian motion; Langevin equation; overdamped equation; Lorentz force; spatially varying; radial symmetry
%}

\maketitle

\section{Introduction}
\label{interoduction}

The Lorentz force due to an external magnetic field modifies the trajectory of a charged, moving particle without performing work on it. This results in characteristic helical trajectories
%of a charged particle 
in case of a constant magnetic field. Such motion is an idealization which compeletely ignores dissipative effects that are highly relevant in, for instance, plasma physics~\cite{goldston1995introduction}. In fact, dissipative effects are dominant in colloidal systems where the dynamics are overdamped. Whereas the effect of Lorentz force in the context of solid-state physics and plasma physics has been throughly studied, much less is known about its effect on diffusion systems subjected to an external magnetic field.

\begin{figure*}
\centering
\resizebox*{0.31\linewidth}{!}{\includegraphics{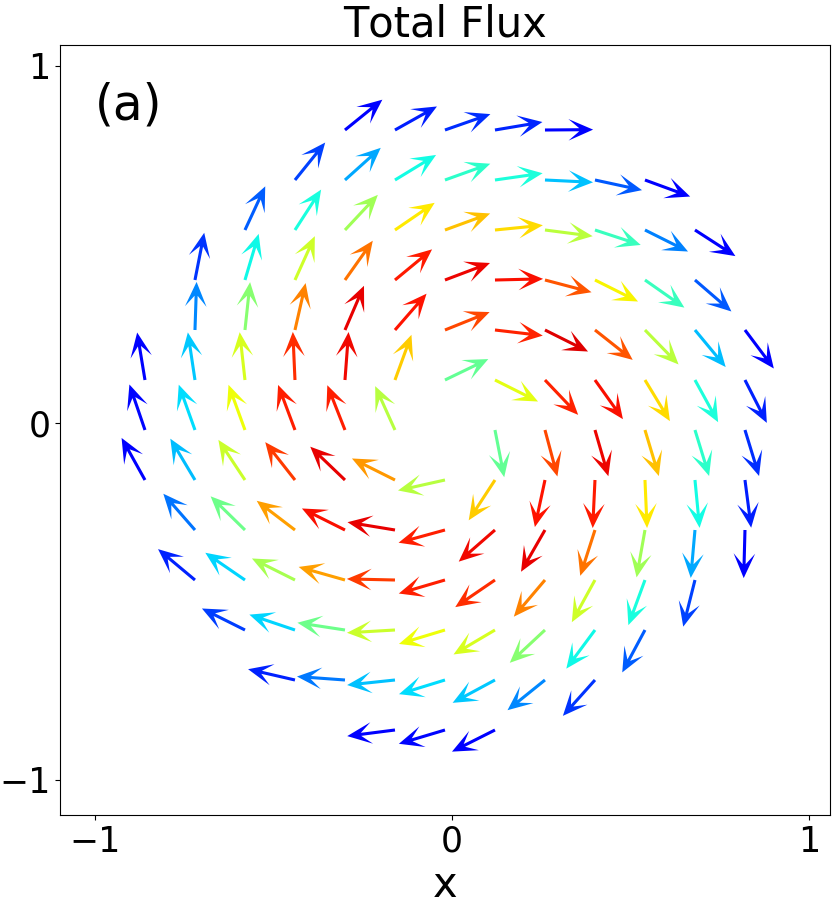}} 
\resizebox*{0.2913\linewidth}{!}{\includegraphics{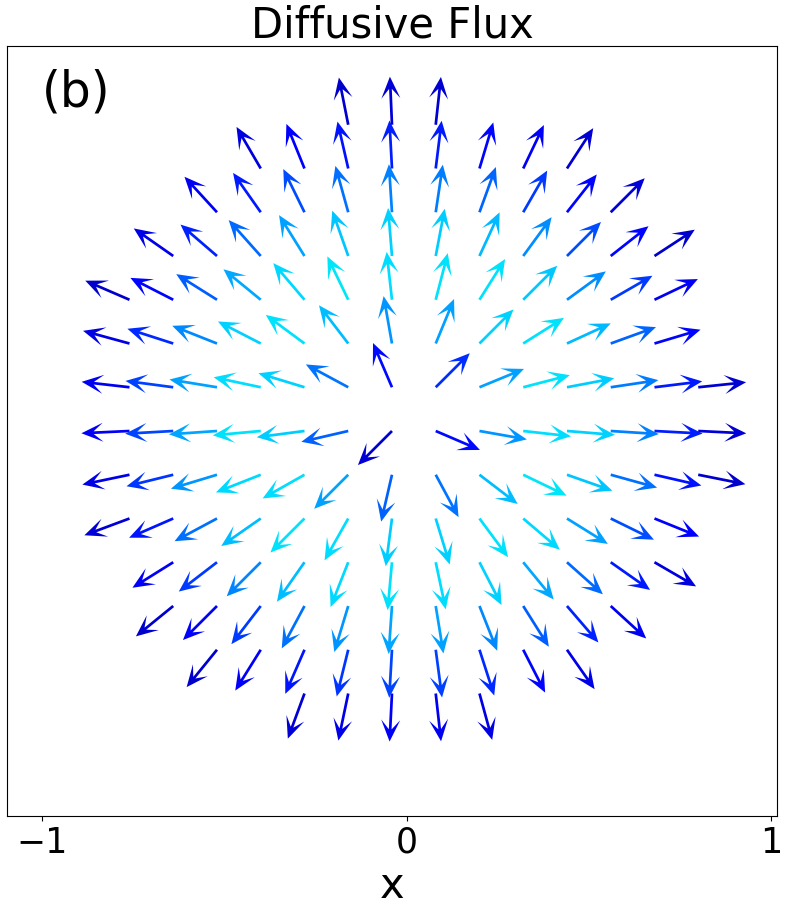}} 
\resizebox*{0.36\linewidth}{!}{\includegraphics{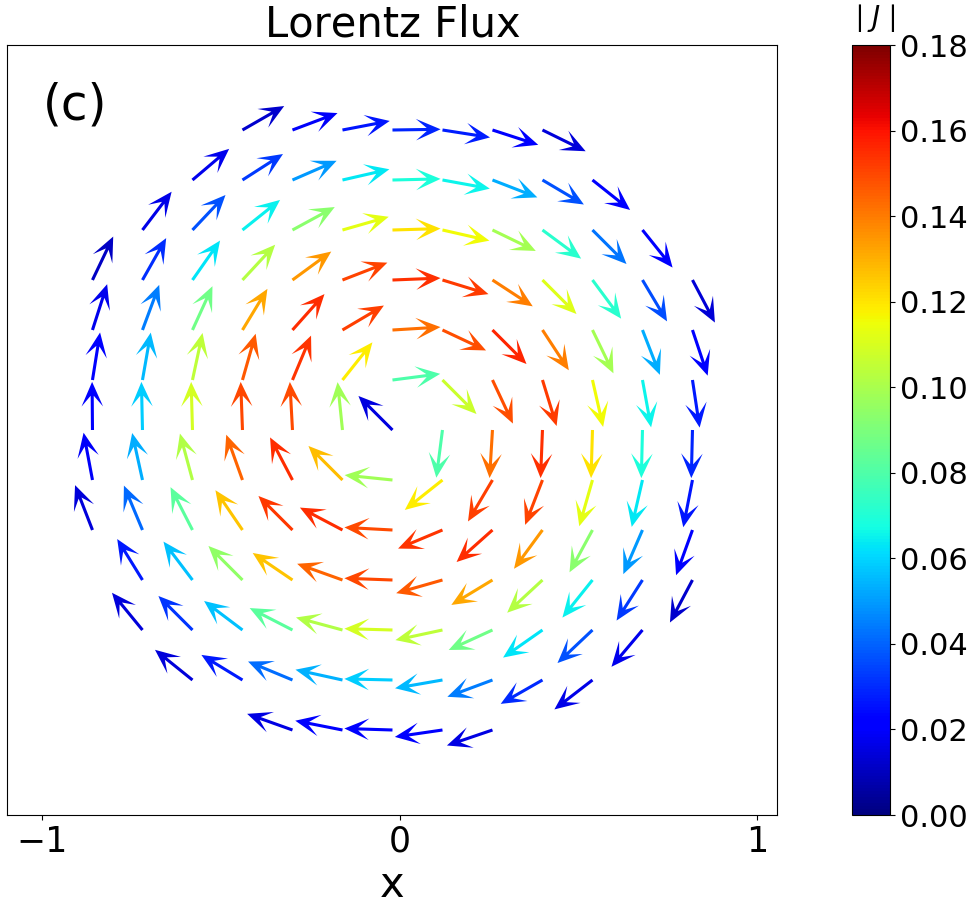}}
\caption{The nonequilibrium dynamics of  a Brownian particle under Lorentz force are different from purely diffusive. A hallmark signature is the appearance of additional Lorentz fluxes which result from the deflection of diffusive fluxes~\cite{abdoli2020nondiffusive}. 
%The particular example here corresponds to \textcolor{blue}{a closed system} with 
The particles are initially distributed in a disc of radius $0.5$ and evolve under Lorentz force due to a constant magnetic field. The figures show the fluxes after $1.0$ Brownian time unit. The total flux, shown in (a), is decomposed in (b) the diffusive flux and (c) the Lorentz flux.
The direction of the fluxes is shown by the arrows; the magnitude is color coded. Note that there is no flux in the steady state of a closed system where density is uniformly distributed
~\cite{abdoli2020nondiffusive}. }
\label{dynamics}
\end{figure*}

A known consequence of the Lorentz force is a reduction of the diffusion coefficient in the plane perpendicular to the magnetic field, whereas the diffusion along the field is unaffected~\cite{balakrishnan2008elements,vuijk2019effect}. The anisotropy in diffusion implies that the corresponding Fokker-Planck equation for the probability distribution acquires a tensorial coefficient, the components of which are determined by the applied magnetic field, the temperature, and the friction coefficient. The tensor, however, is not a typical diffusion tensor due to the antisymmetric elements which account for the fact that Lorentz force curves the trajectory of a charged, diffusing particle, giving rise to additional Lorentz fluxes ~\cite{chun2018emergence,vuijk2019anomalous}. We have recently shown that the dynamics under this tensor are fundamentally different from purely diffusive~\cite{abdoli2020nondiffusive}. In particular, the nonequilibrium dynamics are characerized by features such as additional Lorentz fluxes and a nontrivial density distribution [see Fig.~\ref{dynamics}].
These have implications for dynamical properties of the system such as the mean first-passage time, escape probability, and phase transition dynamics in fluids~\cite{vuijk2019effect,abdoli2020nondiffusive}. 
%*****************************************************************************
%*****************************************************************************

Since the Lorentz force arising from an external magnetic field does no work on the system, the equilibrium properties of the system are unaffected. 
This implies that to observe the nontrivial effects of Lorentz force, the system must be maintained out of equilibrium, possibly in a nonequilibrium steady state. This can be done by driving the system out of equilibrium, for instance, via a time-dependent external potential or shear. Alternatively, one may consider internally driven systems, a particularly interesting example of which is active matter which is ubiquitous in biology~\cite{alvarado2013molecular, alvarado2017force, tan2018self}. We recently demonstrated that a system of active Brownian particles subjected to a spatially inhomogeneous Lorentz force relaxes towards a nonequilibrium steady state with inhomogeneous density distribution and macroscopic fluxes ~\citep{vuijk2019lorenz}. The distinctive dynamics of a charged, passive, diffusing particle under Lorentz force may be appreciated by noting that if the tensor entering the Fokker-Planck equation was positive symmetric, i.e., a diffusion-like tensor, there would be no fluxes in the steady state.

We take a different approach to drive the system into
a nonequilibrium steady state: the particle, while diffusing under the influence of Lorentz force, is stochastically reset to a prescribed location at a constant rate. The concept of stochastic resetting was introduced by Evans and Majumdar~\cite{evans2011diffusion}. In their model, a Brownian particle diffuses freely until it is reset to its initial location. The waiting time between two consecutive resetting events is a random variable for which the Poissonian distribution has been widely used.
Evans and Majumdar showed that diffusion under stochastic resetting gives rise to a nonequilibrium stationary state with a non-Gaussian position distribution and particle flux. They also demonstrated that the mean first-passage time for this model is finite and has a minimum value at an optimal resetting rate. 
Over the last few years, stochastic resetting has been applied to a wide variety of random processes~\cite{evans2011optimal, pal2016diffusion, scacchi2018mean, gupta2019stochastic, pal2019firstE} and generalized to include non-Markovian resetting and dependence of resetting on internal dynamics~\cite{nagar2016diffusion, eule2016non, bodrova2019nonrenewal, falcao2017interacting}. It has been shown that it gives rise to intriguing phenomena such as dynamical phase transitions~\citep{kusmierz2014first, majumdar2015dynamical}, universal properties which are insensitive to details of underlying random process~\cite{reuveni2016optimal, pal2017first, pal2019time} and optimal search strategies~\cite{kusmierz2015optimal}.

In this paper, we show that under stochastic resetting a Brownian system settles into an unusual stationary state which preserves the hallmark features of dynamics under Lorentz force. In the case of a constant magnetic field, the nonequilibrium steady state is characterized by a non-Gaussian probability density, diffusive and Lorentz fluxes. These Lorentz fluxes reflect the behavior shown in Fig.~\ref{dynamics} and are reminiscent of Brownian vortices in a system of colloidal particles diffusing in an optical trap~\cite{roichman2008influence, sun2009brownian, sun2010minimal}. 
Due to the Lorentz force, the flux is not along the density gradient. This holds even for a constant tensorial coefficient. As a consequence, the boundary conditions for diffusion in finite or semi-finite domains take a form different from the typical Neumann or Dirichlet conditions. By considering a minimalistic example, we show how the modified boundary condition gives rise to un unusual stationary state with no counterpart in purely diffusive systems.

The paper is organized as follows. In Sec.~\ref{theory}, we provide a brief theoretical description of diffusion under Lorentz force and stochastic resetting. In Sec.~\ref{NESS}, we derive the steady-state solution to the governing Fokker-Planck equation for constant and inhomogeneous magnetic fields. Finally, we discuss our results and present an outlook in section~\ref{discussion}.

%==========================================================================================
%==========================================================================================

\section{Theory and Simulation}
\label{theory}
%We now present a derivation of our results. 
We consider a single diffusing particle
%a single particle \textcolor{blue}{randomly} diffusing freely 
which is stochastically reset to its initial position $\rr_0$ at a constant rate $\mu$. The particle is subjected to Lorentz force arising from an external magnetic field $\B(\rr)=B(\rr)\nn$ where $\nn$ indicates the direction of the magnetic field and $B(\rr)$ is the magnitude. Our theoretical approach is based on the Fokker-Planck equation for the position distribution of the particle. 
For a spatially inhomogeneous magnetic field, the probability for finding the  particle at position $\rr$ at time $t$, given that it started at $\rr_0$, $p(\rr, t\mid\rr_0)$ obeys the following Fokker-Planck equation~\cite{evans2011diffusion, vuijk2019anomalous, chun2018emergence}
 
\begin{align}
\label{FPE1}
\partial_t p(\rr, t\mid\rr_0) =  & \nabla\cdot[\D(\rr)\nabla p(\rr, t\mid\rr_0)] \\
                                & - \mu p(\rr, t\mid\rr_0) + \mu\delta(\rr-\rr_0), \nonumber
\end{align}
where $\partial_t$ stands for derivative with respect to $t$ and the tensor $\D$ is

\begin{align}
\label{tensord}
\D(\rr) & =  D\left[\left(\boldsymbol{1} + \frac{\kappa^2(\rr)}{1 + \kappa^2(\rr)}\M^2\right) - \frac{\kappa(\rr)}{1+\kappa^2(\rr)} \M \right]  \nonumber \\
        & = \D_s(\rr) + \D_a(\rr), 
\end{align}
where $D = k_B T/\gamma$ is the diffusion coefficient of a freely diffusing particle and $\kappa(\rr) = qB(\rr)/\gamma$ quantifies
%is a parameter quantifying 
the strength of Lorentz force relative to frictional force~\cite{vuijk2019lorenz}. Here $\gamma$ is the friction coefficient, $k_B$ is the Boltzmann constant, $T$ is the temperature and $q$ is the charge of the particle. The matrix $\M$ is defined by $\B(\rr)\times\vv = B(\rr)\M\vv$. $\D_s$ and $\D_a$ are the symmetric and antisymmetric parts of the tensor $\D$.

Note that Eq.\eqref{FPE1} is not of the form of a continuity equation.
The first term on the right hand side of Eq.~\eqref{FPE1} represents the contribution from overdamped motion under Lorentz force. The second and third terms stand for the contribution due to the resetting of the particle: the second term represents the loss of the probability from the position $\rr$ owing to resetting to the initial position $\rr_0$ while the third term stands for the gain of probability at $\rr_0$ due to resetting from all other positions. The flux in the system is given as
\begin{equation}
\label{flux}
\J(\rr, t) = -\D(\rr)\nabla p(\rr, t\mid\rr_0),
\end{equation}
which can be decomposed into the diffusive flux
\begin{equation}
\label{diffusive_flux}
\J_s(\rr, t) = -\D_s(\rr)\nabla p(\rr, t\mid\rr_0),
\end{equation}
and the Lorentz flux
\begin{equation}
\label{Lorentz_flux}
\J_a(\rr, t) = -\D_a(\rr)\nabla p(\rr, t\mid\rr_0).
\end{equation}

Note that the diffusive flux does not depend on the sign of the magnetic field. In contrast, the Lorentz flux can be reversed by reversing the magnetic field. Moreover, it is always perpendicular to the diffusive flux. These properties of Lorentz flux, which are straightforward consequences of how the Lorentz force affects a particle's trajectory, constitute the main rationale behind the above decomposition. Although the dynamics are overdamped it is the presence of these Lorentz fluxes which makes the dynamics under Lorentz force distinct from a purely diffusive system in which only diffusive fluxes exist.

%*****************************************************************************
\begin{figure}
\centering
\resizebox*{1\linewidth}{!}{\includegraphics{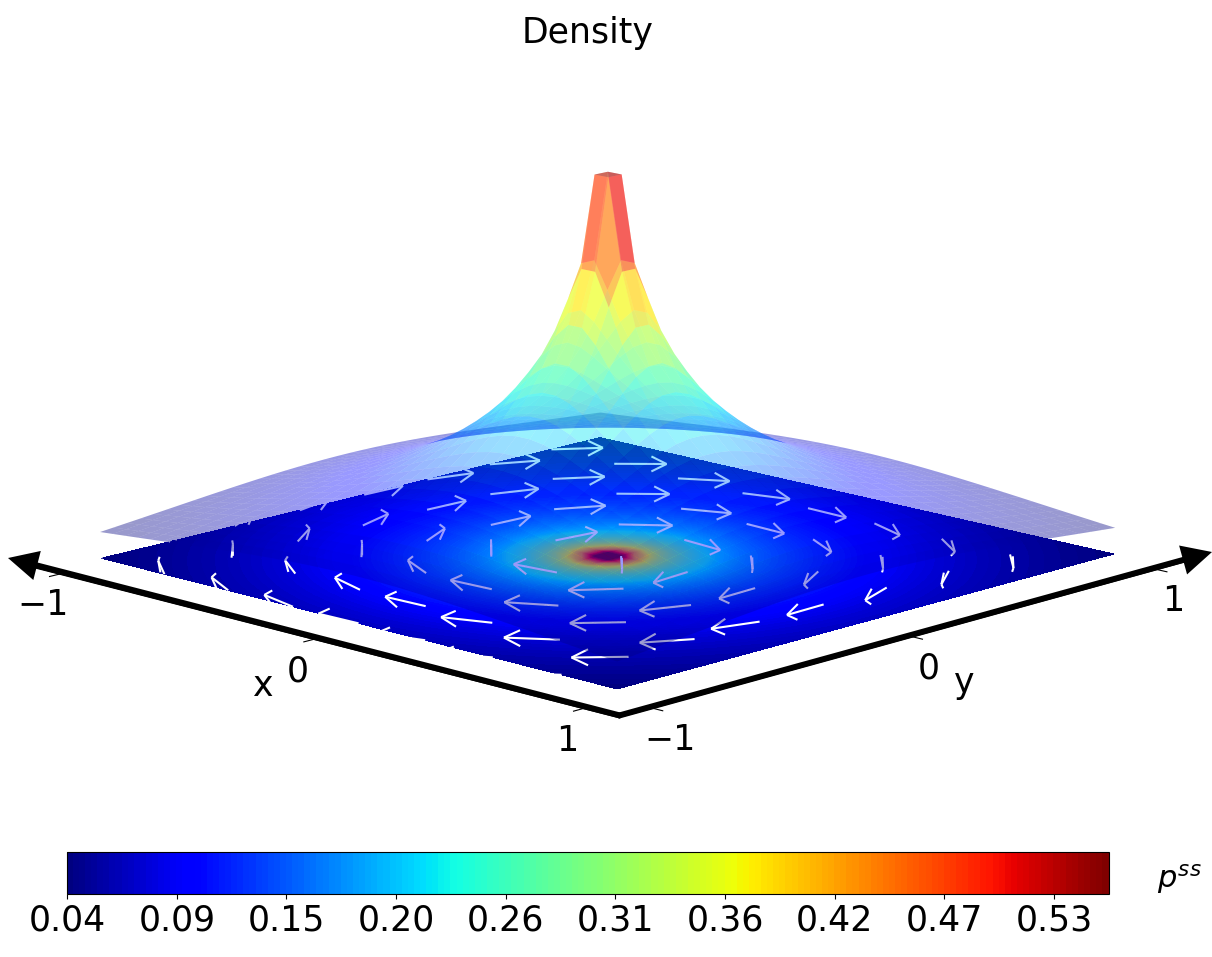}}
\caption{The stationary probability density of the particle's position from Eq.~\eqref{pdf} for a system with $\kappa = 3.0$ is shown in the surface plot on top of the contour plot. %The system size is $2\times 2$. 
The particle is stochastically reset to the origin $\rr_0=\0$ with $\mu = 0.1$. The steady state is characterized by the symmetric, non-Gaussian probability density, the diffusive and Lorentz fluxes. Lorentz fluxes are shown by white arrows.}%The probability density is symmetric and non-Gaussian. The existence of the Lorentz fluxes is shown by white arrows on top of the contour plot.}
\label{density3d}
\end{figure}
%*****************************************************************************
%*****************************************************************************
\begin{figure*}
\centering
\resizebox*{0.3678\linewidth}{!}{\includegraphics{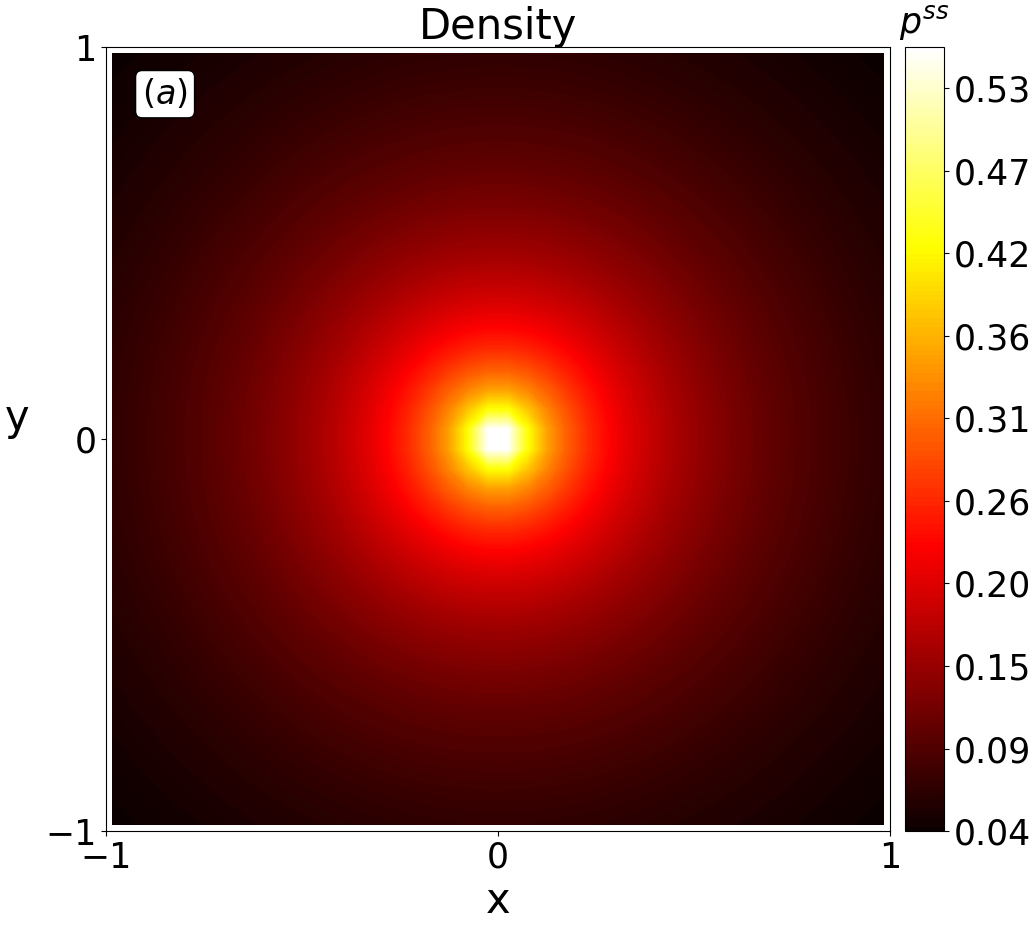}} 
\resizebox*{0.287\linewidth}{!}{\includegraphics{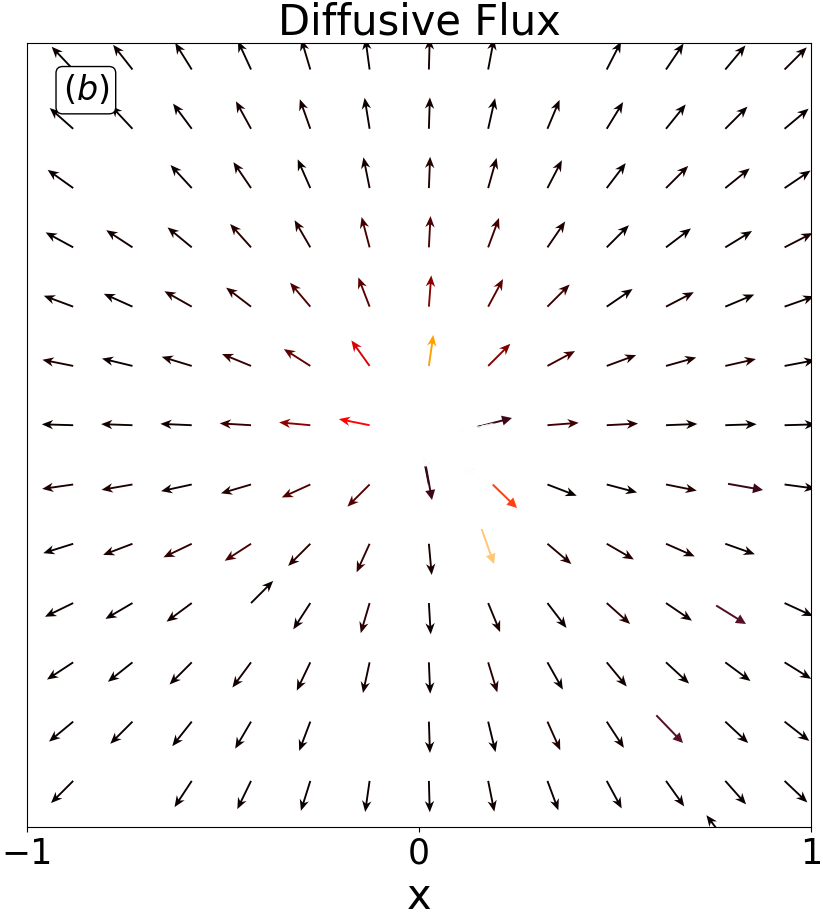}} 
\resizebox*{0.3327\linewidth}{!}{\includegraphics{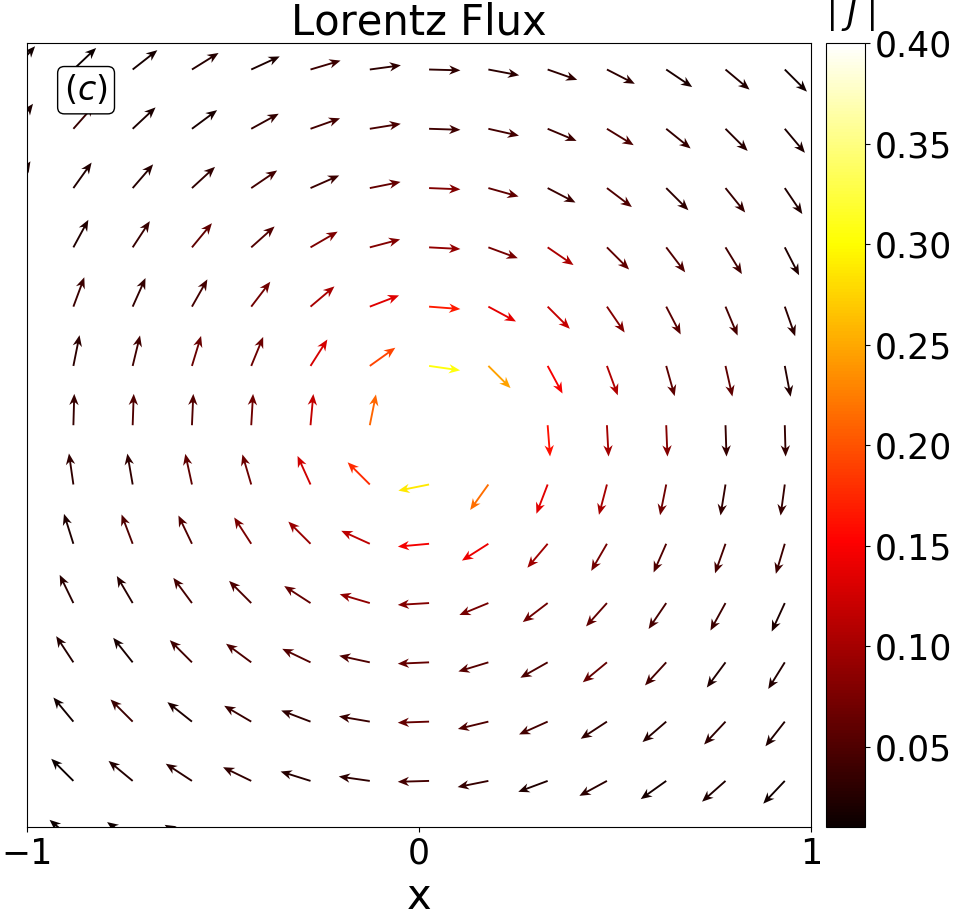}}
\caption{The stationary probability density distribution of the particle's position, the diffusive, and Lorentz fluxes in the system obtained from the underdamped Langevin equation~\eqref{LangevinB} with a mass $m=0.005$ are shown in (a-c), respectively. %The size of the system is $2\times 2$. 
The applied magnetic field is constant such that $\kappa=3.0$. The particle is stochastically reset to its initial position $\rr_0=\0$ at a constant rate $\mu=0.1$. The direction of the fluxes is shown by the arrows; the magnitude is color coded. }
\label{numerics}
\end{figure*}
%*****************************************************************************

In order to confirm our analytical predictions, Brownian dynamics simulations are performed using the Langevin equation of motion~\cite{langevin1908theory}. It has been shown that the overdamped Langevin equation for a Brownian motion in a magnetic field can yield unphysical values for velocity-dependent variables like flux~\citep{vuijk2019anomalous}. Therefore, we use the underdamped Langevin equation with a sufficiently small mass.
Omitting hydrodynamics, the dynamics of the particle are described by the following Langevin equations~\cite{vuijk2019anomalous, abdoli2020nondiffusive}:
  
\begin{align}
\dot \rr(t) &= \vv(t), \nonumber \\
m\dot \vv(t) &= -\gamma \vv + q \vv \times \B(\rr) + \sqrt{2\gamma k_B T}\eeta(t),
\label{LangevinB}
\end{align}
where $m$ is the mass of the particle and $\eeta(t)$ is Gaussian white noise with zero mean and time correlation $\langle\eeta(t)\eeta^{T}(t')\rangle=\boldsymbol{1}\delta(t-t')$. The waiting time between two consecutive resetting events is a random variable with Poisson distribution: in a small time interval $\Delta t$ the particle is either reset to its initial position with probability $\mu\Delta t$ or continues to diffuse with probability $1 - \mu\Delta t$.   
%The correct value of the mass is found out by doing the Brownian dynamics simulation for different values of the mass $m$. 
%\cmnt{As in your previous paper, mention that the magnetic field is applied in the z direction. Because the Lorentz force does not affect the motion in the z direction, we restrict our analysis to the motion in the xy plane. This effectively reduces the problem to two dimensions.}
 Throughout the paper we fix the mass to $m=0.005$ and the integration time step to $dt = 1\times 10^{-6}\tau$ where $\tau=\gamma/k_BT$ is the time for diffusion over one unit distance. %\cmnt{In Ref.~\cite{vuijk2019anomalous} it has been shown that the trajectory of the particle from Eq.~\eqref{LangevinB} converges on that from the right overdamped Langevin equation by decreasing the mass; there is a good agreement for the mass $m=0.02$}. 
In fact, it has been shown that even with a mass $m=0.02$ the trajectory of the particle from Eq.~\eqref{LangevinB} converges on the trajectory from the small-mass limit of this equation~\cite{vuijk2019anomalous}. However, to ensure that the dynamics are overdamped, we have performed simulations with even a smaller mass.
 The simulation results did not show any significant change. Since the magnetic field is applied in the $z$ direction, the Lorentz force has no effect on the motion in this direction. As a consequence, we restrict our analysis to the motion in the $xy$ plane. Accordingly, the vector $\rr$ denotes the coordinates $(x,y)$ of the particle and the tensorial coefficient $\D$ is a $2 \times 2$ matrix.% We consider a square system of size $L^2$, where $L$ is the length in either direction. \cmnt{since we removed the system size from all Figs., shouldn't we remove the last sentence as well?} %The parameter $\kappa$ is varied in the simulations.  

\section{Nonequilibrium steady state}
\label{NESS}
In this section, we determine the steady-state solution to the Eq.~\eqref{FPE1}, first for a constant magnetic field and then for a special choice of spatially inhomogeneous field.
\subsection{Constant magnetic field}
\label{constantmagneticfield}
In the case of a constant magnetic field $\kappa(\rr)=\kappa$, it can be easily shown that $\nabla\cdot[\D_a \nabla p(\rr, t\mid\rr_0)] = 0 $. This implies that the tensor $\D$ in Eq.~\eqref{FPE1} can be replaced by $\D_s = D/(1+\kappa^2)\bold{1}$ which yields the following Fokker-Planck equation:
\begin{align}
\label{FPE}
\partial_t p(\rr, t\mid\rr_0) =  & \frac{D}{1+\kappa^2}\nabla\cdot[\nabla p(\rr, t\mid\rr_0)] \\
                                & - \mu p(\rr, t\mid\rr_0) + \mu\delta(\rr-\rr_0), \nonumber
\end{align}
The steady-state solution $p^{ss}(\rr\mid\rr_0)$ of this equation is obtained by setting $\partial_t p(\rr, t\mid\rr_0)=0$ which, in two dimensions, can be written as~\cite{evans2011diffusion}
\begin{equation}
\label{pdf}
p^{ss}(\rr\mid\rr_0) = \frac{\alpha^2}{2\pi} K_0(\alpha\mid\rr-\rr_0\mid),
\end{equation} 
where $K_0$ is the modified Bessel function of the second kind of order zero and $\alpha = \sqrt{(1+\kappa^2)\mu/D}$. Using Eqs~\eqref{flux} and \eqref{pdf} the diffusive flux can be written as
\begin{align}
\J_{s}(\rr) = \frac{\alpha^3 D}{2\pi(1 + \kappa^2)} K_1(\alpha r)\hat{\rr}, 
\label{radialflux}
\end{align}
where $K_1$ is the modified Bessel function of the second kind of order one, $r = \mid\rr-\rr_0\mid$ is the distance from the starting point of the particle and $\hat{\rr}$ is a unit vector in the radial direction. 

The steady-state solution in case of a constant magnetic field is the same as obtained in Ref.~\cite{evans2011diffusion, evans2014diffusion} with trivial rescaling of the diffusion coefficient wherein $D$ for a freely diffusing particle is replaced by $D/(1+\kappa^2)$ for diffusion under Lorentz force. The distinctive new feature of the steady state is the presence of additional Lorentz fluxes, which can be written as
\begin{align}
\J_{a}(\rr) = -\frac{\alpha^3 D\kappa}{2\pi(1 + \kappa^2)} K_1(\alpha r)\hat{\tet},
\label{circularflux}
\end{align}
where $\hat{\tet}$ is a unit vector in the azimuthal direction. On comparing Eqs.~\eqref{circularflux} and~\eqref{radialflux}, it is evident that the Lorentz flux is nothing else but diffusive flux deflected by the applied magnetic field.% \textcolor{cyan}{My comment: Is this statement really correct? I think it is a component of the deflected flux.}

In Fig.~\ref{density3d} we show a surface plot together with a contour plot of the probability density in the stationary state of the system from Eq.~\eqref{pdf}. The applied magnetic field is such that $\kappa=3.0$. The particle is stochastically reset to its initial position $\rr_0=\0$ at a constant rate $\mu=0.1$. Lorentz fluxes (Eq.~\eqref{circularflux}) are shown as white arrows on top of the contour plot. These fluxes resemble Brownian vortex observed in a system of colloidal particle diffusing in an optical trap~\cite{roichman2008influence, sun2009brownian, sun2010minimal}. Figures~\ref{numerics}(a) to \ref{numerics}(c) show, respectively, the results for the probability density, diffusive fluxes, and the Brownian vortices in the stationary state of the system, obtained from Brownian dynamics simulations. These results are in excellent agreement with the theoretical results shown in Fig.~\ref{density3d}. 

%As shown above, Lorentz fluxes in the steady state result from deflection of the radial fluxes. These Lorentz fluxes slow down the relaxation towards the steady state. However, they do not affect the relaxation dynamics ~\cite{abdoli2020nondiffusive}. This is no longer the case when the magnetic field is inhomogeneous; the \textcolor{blue}{steady-state} solution, as we show below, is determined by the diffusive and Lorentz fluxes.%\cmnt{use this paragraph as the introduction to the next part?}
 %++++++++++++++++++++++++++++++++++++++++++++++++++++++++++++++++++++++++++++++++++++++++
\begin{figure}[b]
%\centering
\begin{center}
%\wspace*{-6mm}
\resizebox*{1\linewidth}{!}{\includegraphics{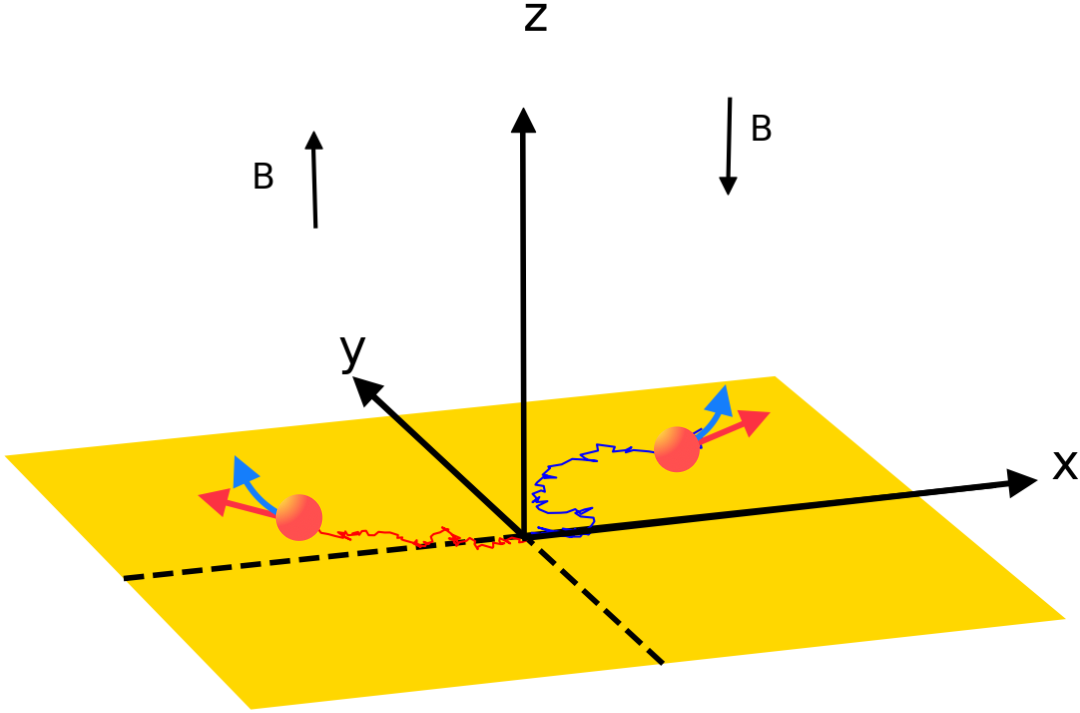}}
%\vspace*{-8mm}
\caption{Schematic showing how the motion of a charged Brownian particle is curved by Lorentz force. Two different trajectories are shown. The particle is subjected to magnetic fields with the same magnitude and opposite directions in each half-plane. The red arrows (straight arrows) depict the direction that the particle follows in the absence of a magnetic field, whereas in the presence of a magnetic field this direction is curved, which is shown in blue.}
\label{schematic02}
\end{center}
\end{figure}
 
%++++++++++++++++++++++++++++++++++++++++++++++++++++++++++++++++++++++++++++++++++++++++++

\subsection{Spatially inhomogeneous magnetic field: a minimal example}
\label{inhomogeneousmagneticfield}
As shown above, Lorentz fluxes in the steady state result from deflection of the radial fluxes. In fact, for a constant magnetic field, they do not affect the relaxation dynamics ~\cite{abdoli2020nondiffusive}. This is no longer the case when the magnetic field is inhomogeneous; the steady-state solution, as we show below, is determined by the diffusive and Lorentz fluxes.

We consider a minimalistic example of a spatially inhomogeneous magnetic field to highlight how the Lorentz fluxes fundamentally alter the boundary conditions giving rise to an unusual stationary state. 
The system is divided into two half-planes by the line $x=0$ [see Fig.~\ref{schematic02}]. Each half plane is subjected to a constant magnetic field with the same magnitude, but opposite direction such that

\begin{equation}
  \kappa(\rr)=\begin{cases}
    -\kappa_0, & \text{$ x \geq 0$},\\
    +\kappa_0, & \text{$ x < 0$}.
  \end{cases}
  \label{magneticfield}
\end{equation}
where $\kappa_0$ is a (constant) parameter. In Fig.~\ref{schematic02}, two different trajectories of the diffusing particle are shown. The red arrows depict the motion of the particle at a given position without Lorentz force, whereas a similar motion in the presence of Lorentz force is shown by blue arrows. As the particle moves away from the origin, the Lorentz force makes the particle undergo a bias toward counterclockwise motion if $x>0$ and clockwise if $x<0$. This implies that there is no flux across the line $x=0$.

This particular choice of the magnetic field ensures that the symmetric part of the tensor, $\D_s$, is a constant tensor in the entire plane, whereas the antisymmetric part, $\D_a$, changes sign at $x=0$. It thus follows that the governing Fokker-Planck equation for the position distribution of the particle is the same as in Sec.~\ref{constantmagneticfield} (Eq.~\eqref{FPE}) with the boundary condition that the $x$ component of the flux (Eq.~\eqref{flux}) is zero at $x=0$. Since the flux is composed of both diffusive and Lorentz components, the boundary condition reads as
\begin{equation}
\label{boundarycondition1}
\ov\cdot\nabla p = 0 \hspace{1cm} \text{at}\hspace{3mm} x = 0,
\end{equation}
where  $\ov = \irow{1,& \kappa}$ is an oblique vector, the direction of which is determined by the magnetic field. This boundary condition is known as the oblique boundary condition and is often employed in theory of wave propagation in presence of obstacles~\citep{gilbarg2015elliptic, keller1981kelvin}. %That the flux in Eq.~\eqref{boundarycondition1} depends on all gradients of the density implies this is a cross-diffusion system. 
Note that for $\kappa = 0$, this reduces to the ordinary Neumann boundary condition. 

The Fokker-Planck equation~\eqref{FPE} with the boundary condition in Eq.~\eqref{boundarycondition1} can be solved using the method of partial Fourier transforms~\citep{hoernig2010green} [see the appendix for details]. The steady-state solution, obtained for $\rr_0 = \0$, is given as
\begin{equation}
\label{inverseFT}
p^{ss}(x, y) = \frac{\alpha^2}{2\pi}\int_0^\infty d\xi\frac{e^{-\beta |x|}}{\beta^2+\kappa^2\xi^2}[\beta \cos(\xi y) - \kappa\xi \sin(\xi y)].
\end{equation}
where $\beta=\sqrt{\xi^2 + \alpha^2}$. %\textcolor{blue}{The analytical evaluation of Eq.~\eqref{inverseFT} is seemingly difficult.}
%It seems difficult to evaluate Eq.~\eqref{inverseFT} analytically. 
One can show that for a system without Lorentz force this expression correctly reduces to the (analytical) results obtained by Evans and Majumdar~\citep{evans2011diffusion}.

%*****************************************************************************
\begin{figure}[t]
\centering
\resizebox*{1\linewidth}{6cm}{\includegraphics{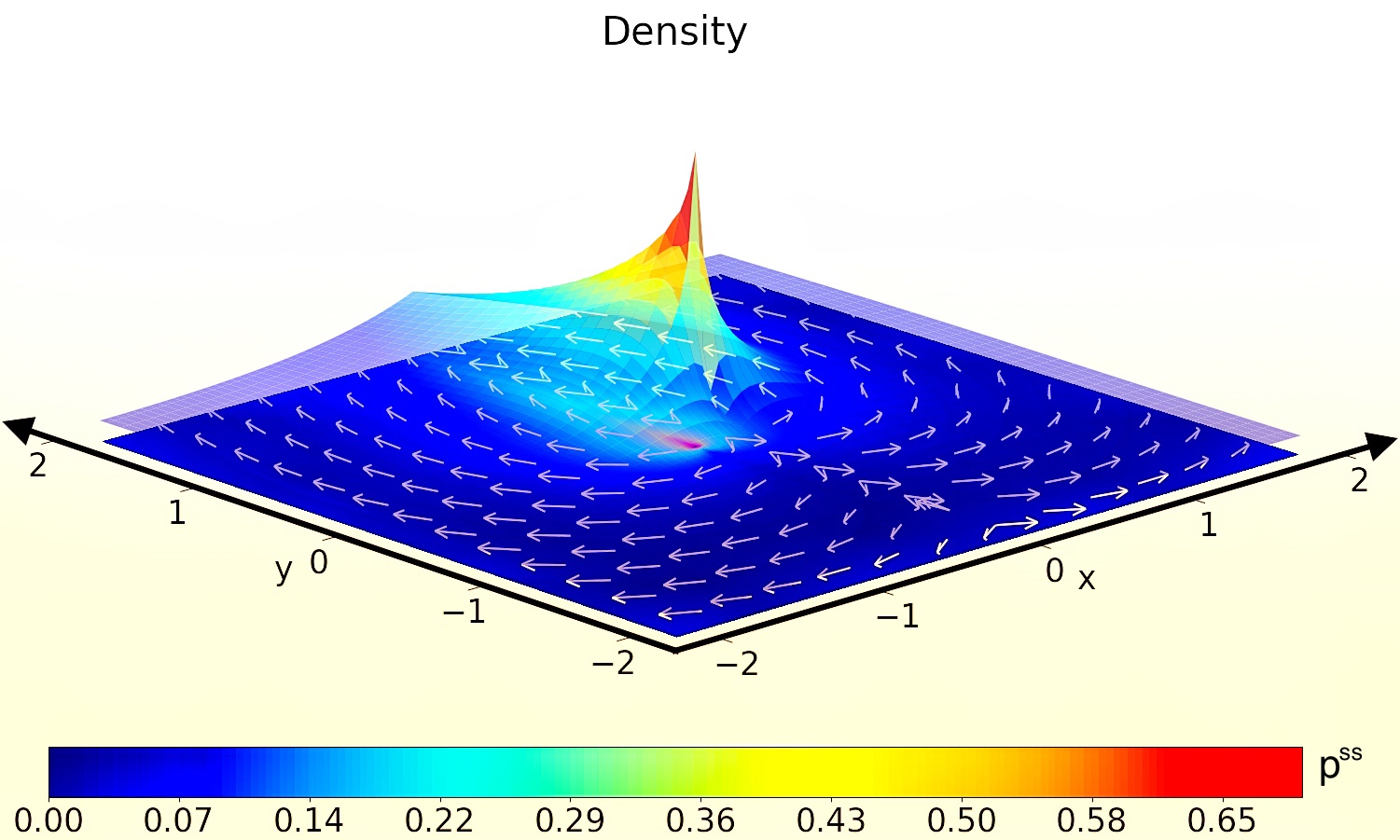}}
\caption{The stationary probability density of the particle's position from Eq.~\eqref{inverseFT} for a system with $\kappa = -2.0$ if $x>0$ and $\kappa=2.0$ otherwise is shown in the surface plot on top of the contour plot. %The system size is $2\times 2$. 
The particle is stochastically reset to the origin $\rr_0=\0$ with $\mu = 0.5$. The steady state is characterized by the non-Gaussian probability density, which is symmetric with respect to the line $x=0$, the diffusive and Lorentz fluxes. Lorentz fluxes are shown by white arrows. }

%The probability density is symmetric with respect to the line $x=0$ and non-Gaussian. The existence of the Lorentz fluxes is shown by white arrows on top of the contour plot.}

\label{density3d2}

\end{figure}
%*****************************************************************************
\begin{figure*}[t]
\begin{center}
\resizebox*{1\linewidth}{!}{\includegraphics{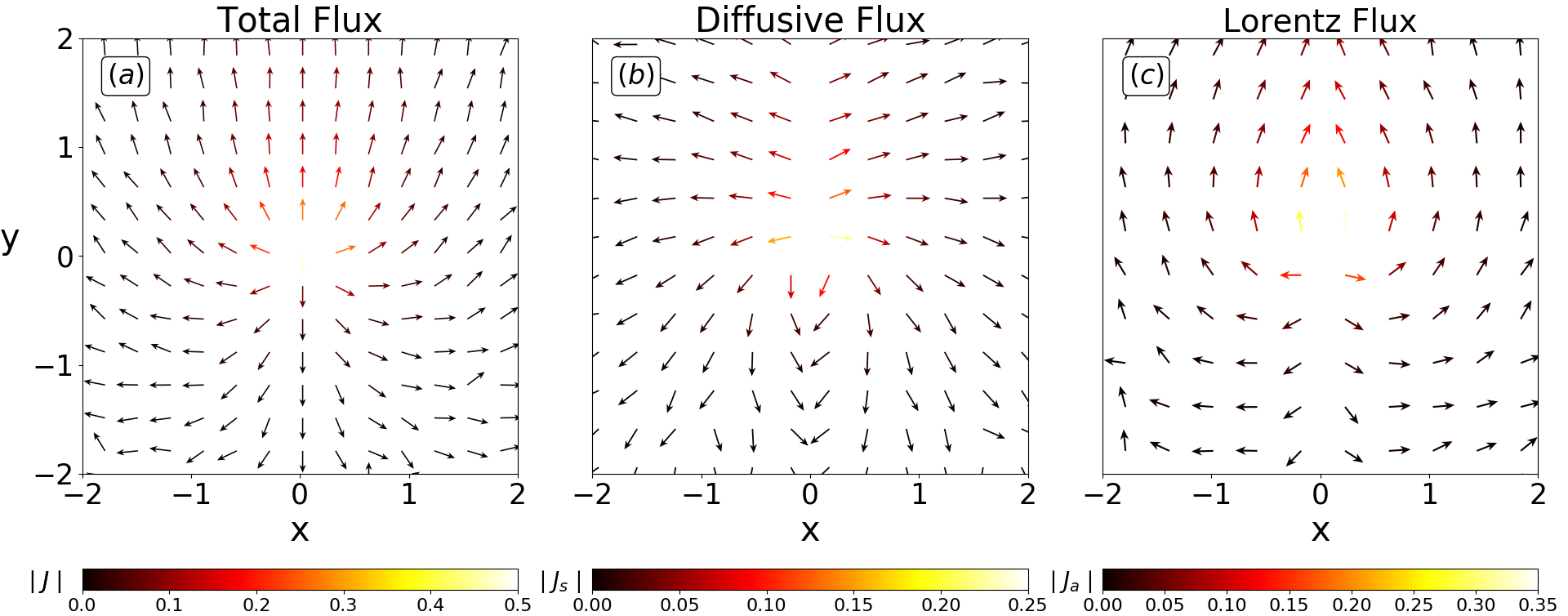}}
%\resizebox*{0.505\linewidth}{!}{\includegraphics{nondiffusiveflux.png}} 
\caption{The total, diffusive, and Lorentz fluxes in the stationary state of a diffusion system subjected to a magnetic field with $\kappa=-2.0$ if $x>0$ and $\kappa=2.0$ otherwise are shown in (a-c), respectively. The particle is stochatically reset to the origin $\rr_0 = \0$ at a constant rate $\mu=0.5$. The results are computed by Brownian dynamics simulations from Eq.~\eqref{LangevinB} with a mass $m=0.005$. The direction of the fluxes is shown by the arrows; the magnitude is color coded. }
\label{fluxes}
\end{center}
\end{figure*}
%++++++++++++++++++++++++++++++++++++++++++++++++++++++++++++++++++++++++++++++++++++++++
%++++++++++++++++++++++++++++++++++++++++++++++++++++++++++++++++++++++++++++++++++++++++
\begin{figure*}
\centering
\resizebox*{1\linewidth}{!}{\includegraphics{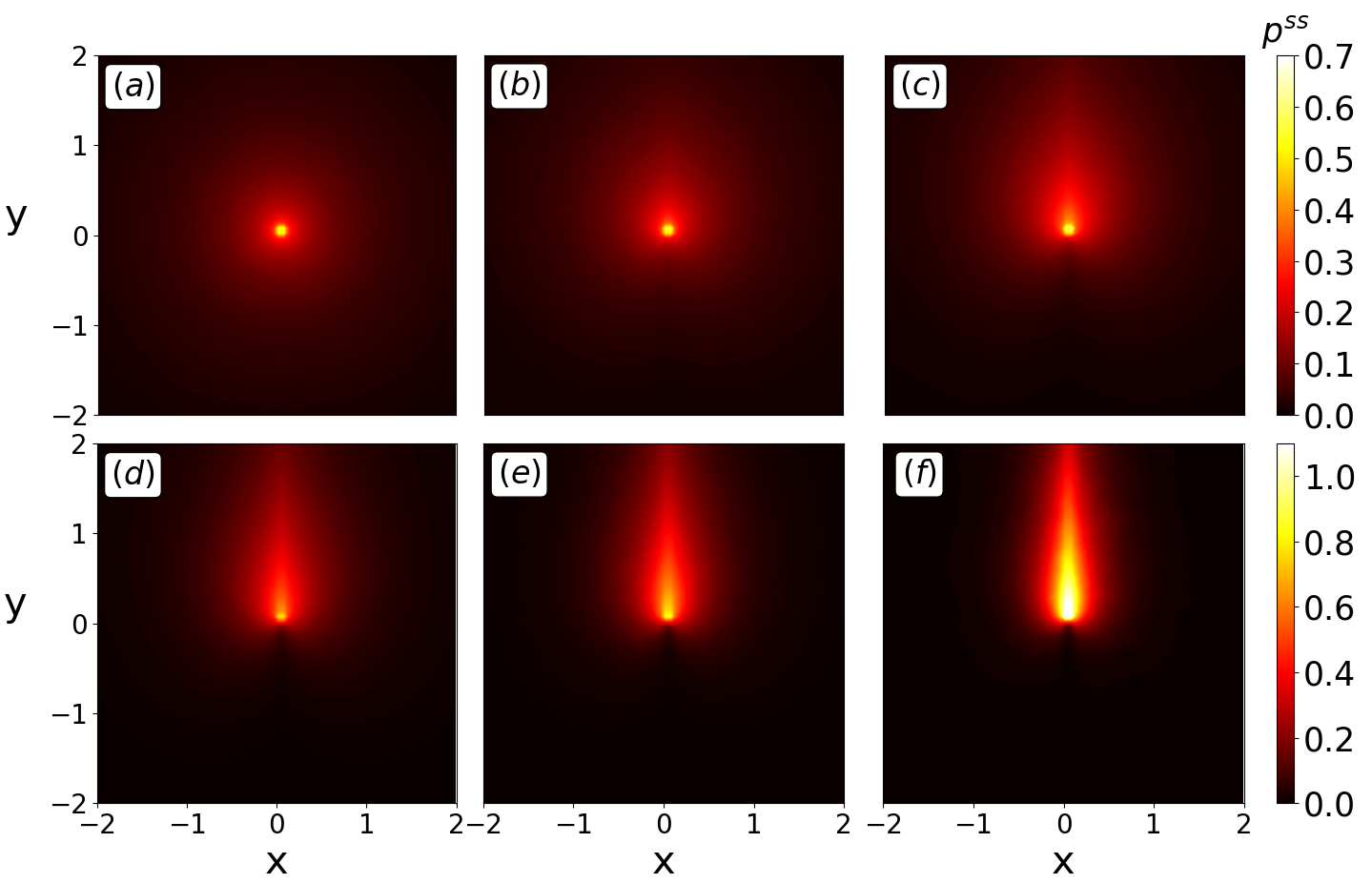}} 
\caption{The stationary probability density distribution of the particle's position. The applied magnetic field is such that $\kappa=-\kappa_0$ if $x>0$ and $\kappa=\kappa_0$ otherwise. $\kappa_0$ is 0.1, 0.5, 1.0, 2.0, 3.0, 5.0 for systems (a) to (f), respectively. The distribution becomes increasingly stretched along the $y$ direction with increasing magnetic field. The results are computed by Brownian dynamics simulations from Eq.~\eqref{LangevinB} with a mass $m=0.005$. The particle is stochastically reset to  the origin $\rr_0=\0$ at a constant rate $\mu=0.5$ in all systems. } 
\label{varyingB}
\end{figure*}
%++++++++++++++++++++++++++++++++++++++++++++++++++++++++++++++++++++++++++++++++++++++++  

%\textcolor{blue}{I broke the figures' paragraph into three paragraphs}
In Fig.~\ref{density3d2} we show a surface plot  on top of a contour plot of the probability density in the stationary state from Eq.~\eqref{inverseFT}. The Lorentz fluxes are shown by white arrows. That an inhomogeneous magnetic field induces an unusual stationary state in the system can be observed by a comparison with Fig.~\ref{density3d}. 

Figure~\ref{fluxes} shows the results from Brownian dynamics simulations with $\kappa_0 = 2.0$ and $\mu=0.5$. The total, diffusive, and Lorentz fluxes in the system are shown in (a-c), respectively. As can be seen in Fig.~\ref{fluxes}, the $x$ component of the total flux is zero at $x=0$. 

Figure~\ref{varyingB} shows the steady-state distribution of the particle's position, obtained from simulations for different values of $\kappa_0$. The particle is stochastically reset to its initial position $\rr_0=\0$ at a constant rate $\mu=0.5$ for all values of $\kappa_0$. As can be seen in Fig.~\ref{varyingB}, the distribution has a candle-flame-like form which is not symmetric with respect to $y$ axis. This can be understood as the accumulation resulting from the equal and opposite Lorentz fluxes at $x=0$. The distribution becomes increasingly stretched along the $y$ direction with increasing magnetic field. A comparison of numerical solutions of Eq.~\eqref{inverseFT} (not shown) with the simulations confirms our analytical predictions. 

\begin{figure*}
\centering
\resizebox*{0.3355\linewidth}{!}{\includegraphics{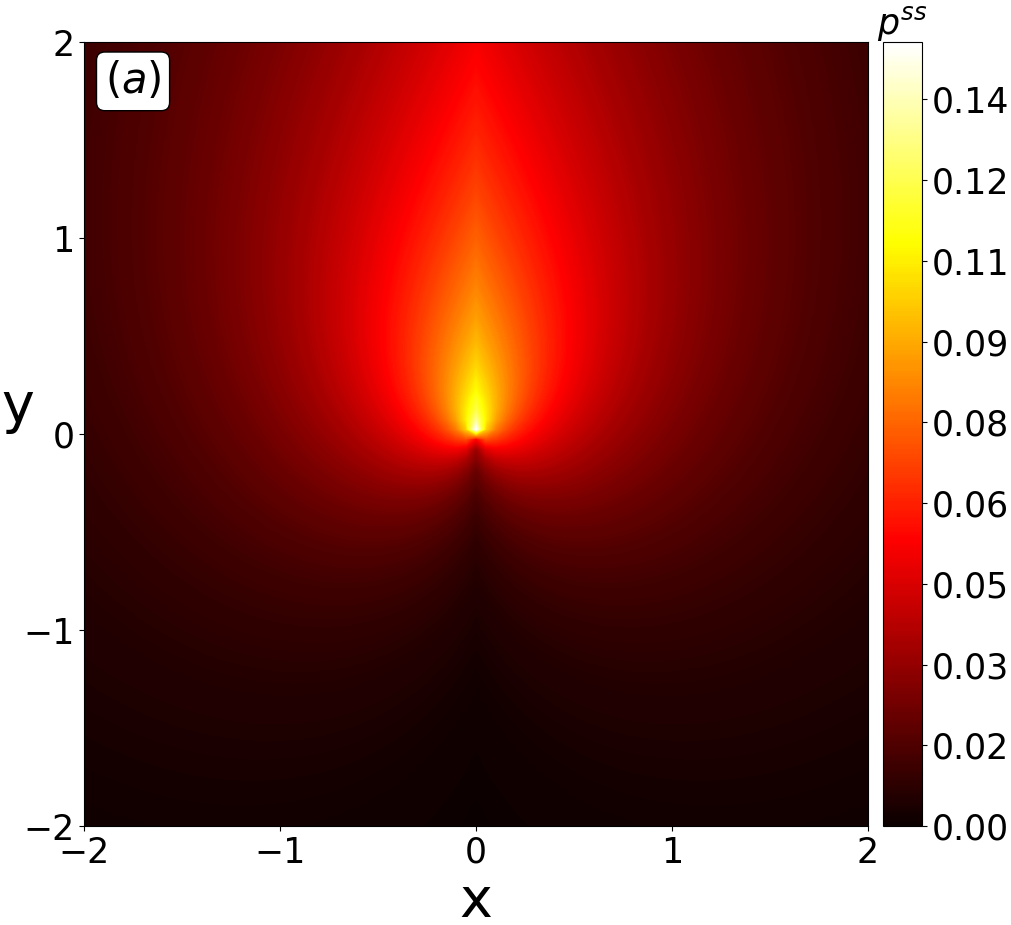}} 
\resizebox*{0.3155\linewidth}{!}{\includegraphics{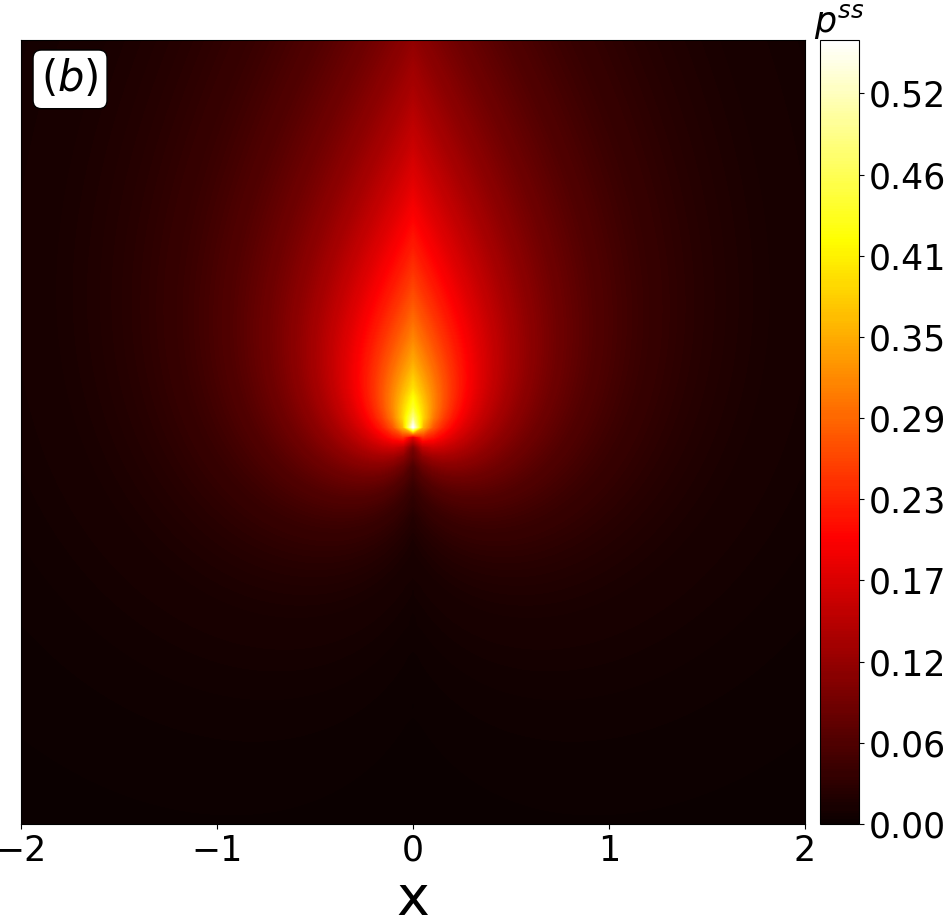}} 
\resizebox*{0.318\linewidth}{!}{\includegraphics{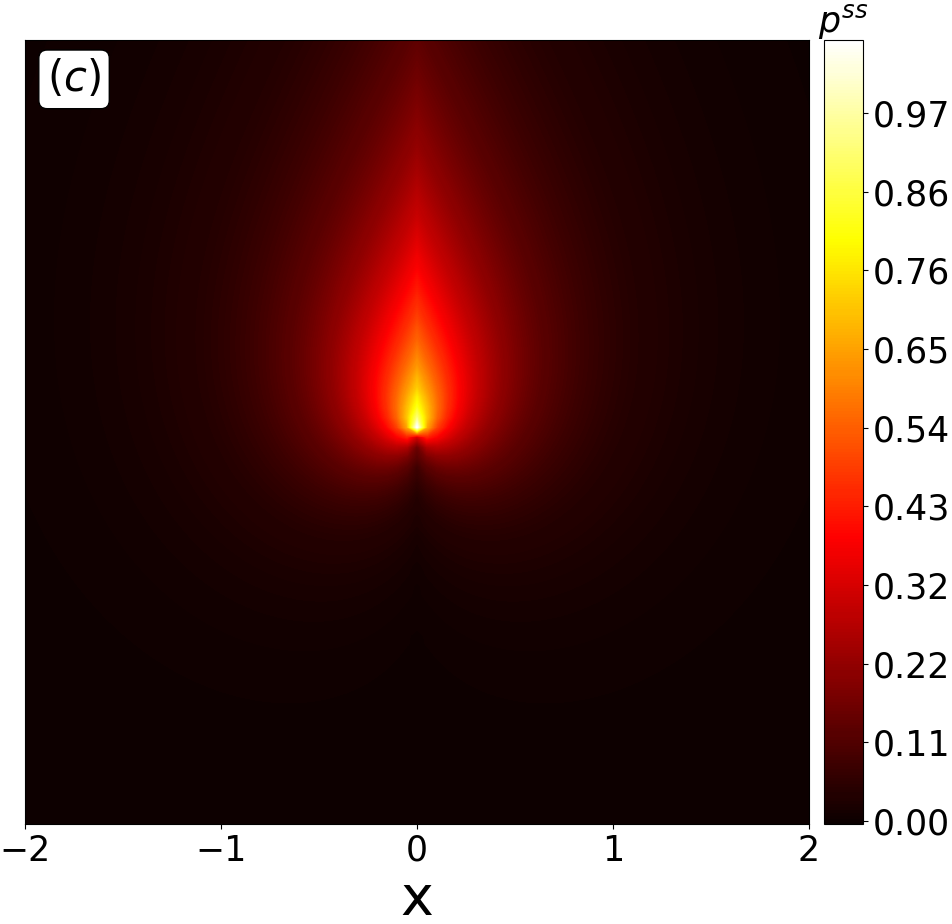}}
\caption{The stationary probability density distribution of the particle's position for different values of $\mu=0.1, 0.4, 0.8$ with $\kappa_0=2.0$ for systems (a) to (c). As in Fig.~\ref{varyingB} the distribution is stretched along the $y$ direction. The width of the distribution along $x$ decreases with increasing $\mu$. %The higher accumulation is resulting from larger rates. 
The results are obtained by Brownian dynamics simulations from Eq.~\eqref{LangevinB} with a mass $m=0.005$.}
\label{varyingrate}
\end{figure*}

In Fig.~\ref{varyingrate} we show the steady-state distribution of the position of the particle from simulations for different values of $\mu= 0.1, 0.4, 0.8$ with $\kappa_0=2.0$. The distribution is stretched along the $y$ direction. The width of the distribution along $x$ direction decreases with increasing $\mu$. 

This minimalistic example shows how Lorentz flux fundamentally alters the probability density and induces an unusual stationary state. Experimentally realizable magnetic fields are likely to have more complicated shapes; however, this does not change the conclusions of this study.%
%*****************************************************************************

%++++++++++++++++++++++++++++++++++++++++++++++++++++++++++++++++++++++++++++++++++++++++++
%==========================================================================================
%==========================================================================================

%\newpage
\section{Discussion and Conclusion}
\label{discussion}

Lorentz force has the unique property that it depends on the velocity of the particle and is always perpendicular to it. Although this force generates particle currents, they are purely rotational and do no work on the system. As a consequence, the equilibrium properties of a Brownian system, for instance the steady-state density distribution, are independent of the applied magnetic field. The dynamics, however, are affected by Lorentz force: the Fokker-Planck equation picks up a tensorial coefficient, which reflects the anisotropy of the particle’s motion. The diffusion rate perpendicular to the direction of the magnetic field decreases with increasing field whereas the rate along the field remains unaffected. In addition to this effect, Lorentz force gives rise to Lorentz fluxes which result from the deflection of diffusive fluxes~\cite{vuijk2019anomalous, abdoli2020nondiffusive}. 

The effects caused by the Lorentz force, however, occur only in nonequilibrium and cease to exist when the distribution of particles reaches equilibrium. A system subjected to stochastic resetting, in contrast, is continuously driven out of equilibrium. In this paper, we showed that by stochastically resetting the Brownian particle to a prescribed location, a nonequilibrium steady state can be created which preserves the hallmark features of dynamics under Lorentz force : a nontrivial density distribution and Lorentz fluxes. We considered a minimalistic example of spatially inhomogeneous magnetic field, which shows how Lorentz fluxes fundamentally alter the boundary conditions giving rise to an unusual stationary state with no counterpart in (purely) diffusive systems. 

One may wonder about the choice of stochastic resetting in this study. 
Although there are several methods to drive a system into a nonequilibrium steady state, stochastic resetting is unique in the sense that it simply renews the underlying (random) process and therefore, in some sense, preserves the dynamics of the underlying process in the steady state. Contrast this with a system of active Brownian particles subjected to Lorentz force~\cite{vuijk2019lorenz} in which Lorentz force couples with the nonequilibrium dynamics of an active particle via its self-propulsion. Although most of the research in stochastic resetting is theoretical, stochastic resetting has been realized experimentally in a system of a colloidal particle which is reset using holographic optical tweezers~\cite{tal2020experimental}. Resetting also features naturally in the measurement of position-dependent diffusion of a particle diffusing near a wall which experiences inhomogeneous drag due to hydrodynamics. The position-dependent diffusion coefficient is measured by letting the particle diffuse freely from a given initial location for a certain period of time before resetting it, using optical tweezers, to the initial location~\citep{dufresne2001brownian}. From the `finite-time' ensemble of measurements, the diffusion coefficient is obtained from the mean squared displacement.

In this work we focused only on the steady-state properties of the system. The investigation of how Lorentz force affects the mean first-passage time and escape probability in such systems is left for a future study.

\section*{Acknowledgments}
We would like to acknowledge Holger Merlitz for fruitful discussions and suggestions.

%*****************************************************************************
%*****************************************************************************  
%\newpage
\appendix
%\appendixpage
%\begin{appendices}
\section{Oblique-Derivative Half-Plane Master Equation}
Partial differential equations with oblique derivative boundary conditions often arise in the theory of waves, for instance, waves on the ocean or in a rotating plane~\citep{gilbarg2015elliptic, keller1981kelvin}. There is a vast amount of mathematical literature on this subject. Here we use the method of partial Fourier transforms adopted from Ref.~\citep{hoernig2010green}. We consider $x = 0$ as a reflecting boundary, for which the zero flux condition can be written as
\begin{equation}
\label{boundarycondition}
\ov\cdot\nabla p = 0 \hspace{1cm} \text{at}\hspace{3mm} x = 0,
\end{equation}
where  $\ov = \irow{1,&\kappa}$ is the oblique vector. We consider a diffusing particle that is stochastically reset to $(x_0, 0)$ at a constant rate $\mu$. Later we will set $x_0 = 0 $ to obtain the solution for our particular case.

The master equation for the stationary probability density $p^{ss}(x, y)$ is
\begin{equation}
\label{stationarymasterequation}
\frac{D}{1+\kappa^2}\nabla^2 p^{ss}(x, y) - \mu p^{ss}(x, y) + \mu\delta(x - x_0)\delta(y) = 0
\end{equation}
 
We define the partial Fourier transform as
\begin{equation}
\label{PFT}
\hat{p}(x, \xi) = \frac{1}{\sqrt{2\pi}}\int_{-\infty}^{\infty} dy\,p^{ss}(x, y) e^{i\xi y} .
\end{equation}
and its inverse as 
\begin{equation}
\label{inversedefinition}
p^{ss}(x, y) = \frac{1}{\sqrt{2\pi}}\int_{-\infty}^\infty d\xi\,\hat{p}(x, \xi)e^{-i\xi y}.
\end{equation}
%\textcolor{blue}{we take the partial Fourier transform of Eq.\eqref{stationarymasterequation}.}
The transformed Fokker-Planck equation [Eq.~\eqref{stationarymasterequation}] becomes
\begin{equation}
\label{PFToME}
\frac{\partial^2\hat{p}(x, \xi)}{\partial x^2} - \beta^2\hat{p}(x, \xi) = -\frac{\alpha^2\delta(x-x_0)}{\sqrt{2\pi}},
\end{equation} 
where $\beta = \sqrt{\xi^2 + \alpha^2}$ and $\alpha = \sqrt{\mu(1+\kappa^2)/D}$. The transformed boundary condition reads as
\begin{equation}
\label{PFToBC}
\frac{\partial\hat{p}(x, \xi)}{\partial x} -i\kappa\xi\hat{p}(x, \xi) = 0 \hspace{1cm} \text{at}\hspace{3mm} x = 0,
\end{equation} 
where $i$ is the imaginary unit. The general solution to Eq.~\eqref{PFToME} is:

\begin{subnumcases}{\hat{p}(x, \xi)=}
Ae^{\beta x} + Be^{-\beta x}, & \text{$0<x<x_0$}, \label{generalsolutiona}\\
Ce^{\beta x} + De^{-\beta x}, & \text{$x>x_0$}.
  \label{generalsolutionb}
\end{subnumcases}

The boundary condition that $\hat{p}(x, \xi)$ is zero as $x \to \infty$ implies $C=0$. That the probability density is continuous on $x=x_0$ implies 
\begin{equation}
\label{continuity}
D = Ae^{2\beta x_0} + B.
\end{equation} 
Substituting Eq.~\eqref{generalsolutiona} into Eq.~\eqref{PFToBC} gives a relationship between $A$ and $B$:
\begin{equation}
\label{AB}
A(\beta -i\kappa\xi) = B(\beta + i\kappa\xi)
\end{equation}

Now one can rewrite Eqs.~\eqref{generalsolutiona} and \eqref{generalsolutionb} as
\begin{equation}
\label{singleconstantsolution}
\hat{p}(x, \xi) = Ae^{\beta x_0}\bigg(e^{-\beta\mid x-x_0\mid} + \Xi e^{-\beta\mid x+x_0\mid}\bigg),
\end{equation}
where $\Xi = (\beta -i\kappa\xi)/(\beta + i\kappa\xi)$. Using this expression one gets%Putting this expression back into Eq.~\eqref{PFToME}, one gets
\begin{align}
\label{secondderivative}
\frac{\partial^2 \hat{p}(x, \xi)}{\partial x^2} = Ae^{\beta x_0}\big[ & \beta^2e^{-\beta\mid x-x_0\mid}-2\beta\delta(x-x_0)e^{-\beta\mid x-x_0\mid} \nonumber \\
& + \Xi\beta^2 e^{-\beta\mid x+x_0\mid} \big]. 
\end{align}
The second derivative of $\hat{p}(x, \xi)$ in Eq.~\eqref{PFToME} can be replaced by 
Eq.~\eqref{secondderivative}, which results in $A = (\alpha^2e^{-\beta x_0})/(2\beta\sqrt{2\pi})$. After some simplifications one gets
\begin{equation}
\label{simplifiedME}
\hat{p}(x, \xi) = \frac{\alpha^2}{\sqrt{2\pi}}\bigg(\frac{e^{-\beta\mid x-x_0\mid}-e^{-\beta\mid x+x_0\mid}}{2\beta} +\frac{e^{-\beta\mid x+x_0\mid}}{\beta+i\kappa\xi}\bigg).
\end{equation}  
For the system studied in this paper, we set $x_0 = 0$. Thus
\begin{equation}
\label{limitedsolution}
\hat{p}(x, \xi) = \frac{\alpha^2}{\sqrt{2\pi}}\frac{e^{-\beta x}}{\beta + i\kappa\xi}.
\end{equation}
We could not find a closed analytical form for the inverse Fourier transform of Eq.~\eqref{limitedsolution}. Nevertheless, the following intergal can be evaluated numerically to obtain the steady-state solution:
\begin{equation}
\label{IFT}
p^{ss}(x, y) = \frac{\alpha^2}{2\pi}\int_0^\infty d\xi\frac{e^{-\beta |x|}}{\beta^2+\kappa^2\xi^2}[\beta \cos(\xi y) - \kappa\xi \sin(\xi y)].
\end{equation}
Note the factor $1/2$ on the right hand side of the Eq.~\eqref{IFT}, which accounts for the (symmetric) extension of the solution to the $x<0$ half-plane. 
%As a check, it can be easily shown that for
For special case of $\kappa = 0$, it is easy to show that the above integral reduces to 
\begin{equation}
p^{ss}(\rr) = \frac{\alpha^2_0}{2\pi}K_0(\alpha_0 |\rr|),
\end{equation}
where $\alpha_0 = \sqrt{\mu/D}$ where $|\rr|$ is the distance from the origin, same as reported in Ref.~\citep{evans2011diffusion} for a two dimensional (symmetric) diffusion under stochastic resetting.
%\end{appendices}

\newpage
\newpage
\providecommand{\noopsort}[1]{}\providecommand{\singleletter}[1]{#1}%

\end{document}